\documentclass[lettersize,journal]{IEEEtran}
\usepackage{amsmath,amsfonts}
\usepackage{algorithmic}
\usepackage{algorithm}
\usepackage{array}
\usepackage[caption=false,font=normalsize,labelfont=sf,textfont=sf]{subfig}
\usepackage{textcomp}
\usepackage{stfloats}
\usepackage{url}
\usepackage{verbatim}
\usepackage{graphicx}
\usepackage{cite}
\usepackage{tabularx}
\usepackage{booktabs}
\usepackage{multirow}
\usepackage{gensymb}
\usepackage{longtable}
\usepackage{enumitem}

\usepackage{hyperref}
\hypersetup{breaklinks=true}
\begin{document}

\title{Advanced Capacity Accreditation of Future Energy System Resources with Deep Uncertainties}

\author{Ethan~Cantor,~\IEEEmembership{Student Member,~IEEE,}
        Yinyin~Ge,~\IEEEmembership{Member, IEEE,}
        Hongxing~Ye,~\IEEEmembership{Senior Member, IEEE,}
        Jie~Li,~\IEEEmembership{Senior Member, IEEE,}
}
\IEEEaftertitletext{\vspace{-20pt}}

\maketitle

\begin{abstract}
The electric power sector has seen an increased penetration of renewable energy sources (RESs) that could strain the system reliability due to their inherent uncertainties in availability and controllability. Effective load carrying capability (ELCC) is widely used to quantify the reliability contributions of these RESs. However, existing ELCC methods can over- or under-estimate their contributions and often neglect or simplify other critical factors such as transmission constraints and evolving climate trends, leading to inaccurate capacity credit (CC) allocations and inefficient reliability procurement in capacity markets. To address these limitations, this paper proposes TRACED (TRansmission And Climate Enhanced Delta)--an advanced capacity accreditation approach that integrates transmission constraints and climate-adjusted system conditions into a Delta ELCC evaluation. Case studies on a modified IEEE-118 bus system with high RES and energy storage penetrations demonstrate that TRACED produces portfolio-consistent CC allocations by capturing resource interactions and avoiding the double-counting of shared reliability benefits inherent in marginal ELCC, which may otherwise lead to under-procurement of reliability resources. Results further demonstrate that transmission congestion and evolving climate trends have mutual impacts on CC allocation, justifying their necessary integration into TRACED. 
\end{abstract} 

\vspace{-8pt}
\section*{Nomenclature}
\vspace{-3pt}
\noindent\textbf{Sets and Indices}
\noindent
\begin{description}[
  leftmargin=1.6cm,
  labelwidth=1.4cm,
  labelsep=0.2cm,
  align=left,
  itemsep=0pt,
  topsep=2pt
]
  \item[$\mathcal{B}$] Set of battery storage systems, indexed by $b$.
  \item[$\mathcal{H}$] Set of historical data years, indexed by $h$.
  \item[$\mathcal{I}$] Set of thermal generators, indexed by $i$.
  \item[$\mathcal{J}$] Set of generators considered for capacity accreditation, indexed by $j$.
  \item[$\mathcal{L}$] Set of transmission lines, indexed by $l$.
  \item[$\mathcal{N}$] Set of transmission nodes, indexed by $n$.
  \item[$\mathcal{S}$] Set of solar photovoltaic generators, indexed by $s$.
  \item[$\mathcal{T}$] Set of time steps, indexed by $t$.
  \item[$\mathcal{W}$] Set of wind generators, indexed by $w$.
\end{description}

\vspace{3mm}
\noindent\textbf{Decision Variables}

\begin{description}[
  leftmargin=1.6cm,
  labelwidth=1.4cm,
  labelsep=0.2cm,
  align=left,
  itemsep=0pt,
  topsep=2pt
]
  \item[$CH_{b,t}$] Energy charged by battery $b$ at time $t$ [MW].
  \item[$DIS_{b,t}$] Energy discharged by battery $b$ at time $t$ [MW].
  \item[$ch_{b,t}$] Charging on/off state of battery $b$ at time $t$.
  \item[$curt_{s/w,t}$] Electric energy curtailment of solar generator $s$/wind generator $w$ at time $t$ [MW].
  \item[$dis_{b,t}$] Discharging on/off status of battery $b$ at time $t$.
  \item[$G_{i/s/w,t}$] Electric energy generation at time $t$ of thermal generator $i$/solar generator $s$/wind generator $w$ [MW].
  \item[$LS^{sys}_{t}$] System load shedding at time $t$ [MW].
  \item[$LS_{n,t}$] Load shedding of node $n$ at time $t$ [MW].
  \item[$ls_{t}$] System load shedding on/off status at time $t$.
  \item[$SOC_{b,t}$] State of charge of battery $b$ at end of time $t$ [\%].
  \item[$sd/su_{i,t}$] Shut-down/start-up  action of thermal generator $i$ at start of time $t$.
  \item[$u_{i,t}$] On/off status of thermal generator $i$ at time $t$.
\end{description}

\vspace{3pt}
\noindent\textbf{Parameters}

\begin{description}[
  leftmargin=1.6cm,
  labelwidth=1.4cm,
  labelsep=0.2cm,
  align=left,
  itemsep=0pt,
  topsep=2pt
]
  \item[$AAR_{l, \tau_{\text{air}}}$] Temperature derating coefficient of line $l$ at air temperature $\tau_{\text{air}}$ [\%/\degree C].
  \item[$C_{b/i/s/w}$] Cost of generation of battery $b$/thermal generator $i$/solar generator $s$/wind generator $w$ [\$/MW].
  \item[$C^{curt}_{s/w}$] Cost of curtailment of solar generator $s$/wind generator $w$ [\$/MW].
  \item[$D_{n,t}$] Energy demand of node $n$ at time $t$ [MW].
  \item[$DT_{i}$] Minimum downtime of thermal generator $i$.
  \item [$E^{max}_b$] Maximum capacity of battery $b$ [MW].
  \item[$FI_j , LI_j$] First-in/last-in CCs of generator $j$ [MW].
  \item[$FOR_{i,\tau_{\text{air}}}$] Forced outage rate of thermal generator $i$ at air temperature $\tau_{\text{air}}$.
 \item[$G^{min\hspace{-0.5mm}/\hspace{-0.5mm}max}_i$] \hspace{-0.2mm}Minimum/maximum power output of thermal generator $i$ [MW].
  \item[$\gamma_{s/w}$] Nominal power of solar farm $s$/wind farm $w$ [MW].
  \item[$\eta^{ch/dis}_{b}$] Charging/discharging efficiency of battery $b$.
  \item[$\eta_{s/w}$] Efficiency of solar generator $s$/wind generator $w$.
  \item[$PTDF_{n,l}$] Power transmission distribution factor of node $n$ to line $l$.
  \item[$SUC_{i}$] Start-up cost of thermal generator $i$ [\$].
  \item[$SDC_{i}$] Shut-down cost of thermal generator $i$ [\$].
  \item [$SOC^{min}_b$] Minimum state of charge of battery $b$ [\%].
  \item [$SOC^{max}_b$] Maximum state of charge of battery $b$ [\%].
  \item[$UT_{i}$] Minimum uptime of thermal generator $i$.
  \item[$VOLL$] Value of Lost Load [\$/MW].
  \item[$\tau_{\text{air,t}}$] Air temperature at time $t$ [\degree C].
\end{description}

\vspace{-5pt}
\section{Introduction}

\IEEEPARstart{T}{he} U.S. electric power sector is rapidly transitioning towards increased integration of renewable energy sources (RESs), primarily solar and wind, paired with energy storage. This shift supports a sustainable and diversified energy portfolio with enhanced energy security. While providing low-cost, carbon-free generation, RESs introduce new operational challenges to the grid, especially in maintaining reliability under increasingly frequent and severe extreme weather events.

Unlike conventional dispatchable generators, most RESs lack controllability, making their reliability contribution difficult to quantify, particularly when peak demand does not coincide with peak RES outputs. As RES penetrations increase, such reliability uncertainty leads to significant concerns, elevating the risks of component and system failures. While this penetration threshold varies by system, a commonly cited benchmark is 30\%~\cite{MISO RIIA}. For example, MISO’s Renewable Integration Impact Assessment reports significant reliability challenges beyond 30\% RES penetration without transformative system changes~\cite{MISO RIIA}, with similar findings for larger regions requiring substantial flexibility enhancements \cite{M. Milligan 2015}. As the nation's RES capacity continues to grow, accurately quantifying their reliability contribution becomes essential for planning and investment decisions and North American Electric Reliability Corporation (NERC) compliance. Such models are critical to maintaining a reliable electricity supply while enabling continued RES integration.

Reliability of bulk energy systems is defined as their ability to deliver uninterrupted electricity to customers \cite{Allan and Billinton 2020}. The most widely adopted metric is the Loss of Load Expectation (LOLE)--the expected number of days per year when supply fails to meet demand \cite{Allan and Billinton 2020}. NERC Standard BAL-502-RF-03 \cite{BAL-502-RF-03} requires Regional Transmission Organizations (RTOs) and Independent System Operators (ISOs) to establish resource adequacy criteria, commonly adopting $\leq$0.1 days/year. Others have adopted a Loss of Load Hours (LOLH) of $\leq$2.4 hours/year to capture multiple outages per day \cite{PNNL Kintner-Meyer}. To meet reliability requirements, generators are assigned reliability contributions known as capacity credits (CCs), a portion of their nameplate capacities that can be relied upon during high-demand periods. Traditional fossil-fired units typically receive CCs near 100\% of their capacities due to controllability, whereas RESs often receive significantly lower CCs \cite{PJM ELCC Class Ratings}. CCs determine generator compensation in capacity markets, where RTOs/ISOs contractually purchase capacity for high-demand periods, incentivizing generators to improve dependable capacity during system stress.

Various methods have been developed to calculate CCs of different types of energy resources, generally categorized as approximation and reliability methods. Approximation methods are used by ISOs/RTOs for capacity planning due to lower data requirements and computational simplicity, but they often produce less accurate CC estimates, particularly in systems with high RES penetrations \cite{S.H. Madaeni}. Most ISOs/RTOs have implemented reliability-based methods, the most common being the effective load carrying capability (ELCC), describing the additional load a new generator can support to maintain the same system reliability level \cite{Garver 1966}.

Variants of the ELCC method include the average and marginal ELCC. Average ELCC assigns a uniform CC to all resources within a class based on their aggregate contribution to system adequacy, effectively treating the class contribution as static to simplify computation. However, it cannot capture diminishing reliability contributions at high penetrations, often overestimating later-added RES CCs and risking under-procurement. Differently, marginal ELCC calculates the incremental contribution of new resources, accounting for their interactions with existing resources. For example, wind plants with correlated outputs can shift net load peaks and reduce the marginal reliability value of additional wind capacity. By reflecting these interactions, marginal ELCC provides stronger signals for complementary resource investment, leading to improved market efficiency \cite{A.Pham NREL, Aagaard and Kleit 2023}. PJM and NYISO have transitioned from average to marginal ELCC in 2024 \cite{PJM Capacity Market Reform, NYISO Capacity Market Reform}, and ISONE is planning for the adoption in 2028 \cite{ISONE Capacity Market Reform}. However, marginal ELCC often underestimate RESs' contributions, with the sum of individual CCs less than the portfolio total, leading to extraneous consumer costs~\cite{Aagaard and Kleit 2023}. 

Two approaches may advance the marginal ELCC method: the Delta method and the Integration method. The Delta method \cite{N. Schlag} ensures that the sum of individual generator CCs equals the portfolio CC ($PORT$) while considering the difference between its first-in (FI) ELCC (i.e., marginal ELCC of a generator added first to the system) and last-in (LI) ELCC (i.e., marginal ELCC of that generator added last to the system). The portfolio interactive effect (PIE)--the difference between the portfolio ELCC and the sum of LI ELCCs--is proportionately distributed based on each generator's individual interactive effect (IIE), defined as the difference between its FI and LI ELCCs. The Integration method \cite{K. Carden} further refines this concept by integrating across multiple installation levels between the FI and LI points, capturing the non-linear behavior of the ELCC curve. The Integration method is typically useful at the class level, losing much of individual generators' details. The FI, LI, Delta, and Integration ELCCs are illustrated in Figure \ref{Marginal, Delta, and Integration ELCC Graphic}. Neither method, however, considers the impact of generator location in the system.  

\begin{figure}[h]
\vspace{-10pt}
\centering
\includegraphics[width=1.7in]{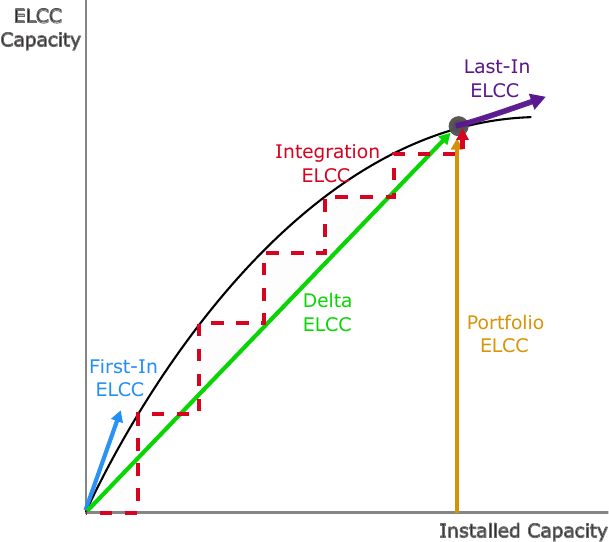}
\vspace{-6pt}
\caption{Marginal, Delta, and Integration ELCC methods.}
\vspace{-5pt}
\label{Marginal, Delta, and Integration ELCC Graphic}
\end{figure}

Indeed, a generator's location affects its reliability contribution in two dimensions. First, locational transmission capacity may limit the ability to deliver power to loads, reducing the generator CC. However, current practice typically separates transmission constraints from ELCC calculations. For example, PJM uses the Capacity Transfer Emergency Objective \cite{PJM Manual 20} to calculate LOLE specifically due to inter-load serving area (LSA) transmission constraints, but intra-LSA limits are ignored and generator CCs remain unaffected, reducing reliability evaluation versatility. Research has explored incorporating transmission limits into ELCC calculations. Convexified unit commitment (UC) constraints were used to better assess CCs of wind, solar, and energy storage resources \cite{S. Wang 2022}, but only considered an oversimplified two-node transmission system. A security-constrained UC problem was studied to solve the ELCC of wind and solar resources  \cite{Z. Chen 2015 Conference}, and tested on the IEEE-118 bus system to demonstrate its applicability to RTO/ISO scale systems \cite{Z. Chen 2015 Journal}. A transmission-constrained UC problem was applied to the Chilean Central Interconnected System, using the PLEXOS software to evaluate wind CCs \cite{R. Wetzel 2014}. These works used either the average or marginal ELCC, inheriting their respective limitations.

Moreover, locational weather conditions strongly affect generator performance, particularly for RESs, impacting their reliability contributions. As such, RTOs/ISOs have incorporated forecasted weather into ELCC calculations, but assume that future weather will resemble historic patterns, ignoring evolving weather trends \cite{PJM Manual 20, J. Black and V. Rojo}. This assumption could underestimate their impacts, especially as extreme weather events become more frequent and severe. 

This work addresses the above limitations by incorporating transmission constraints and evolving weather patterns for more realistic ELCC assessments. The major contributions are:

\begin{enumerate}
    \item A new ELCC method, TRACED (TRansmission And Climate Enhanced Delta), is proposed to provide portfolio-consistent CCs while incorporating key system factors that influence reliability contributions. Specifically, TRACED introduces capacity accreditation corrections in both spatial and temporal dimensions, addressing transmission constraints and deep uncertainties associated with RES outputs against evolving climate conditions.

    \item A transmission-constrained UC framework is developed to evaluate ELCC, with detailed component modeling and explicit modeling of network limitations as a spatial correction to capacity accreditation. The computational burden is mitigated through a reduced-complexity formulation and a rolling-horizon solution approach.

    \item Evolving weather impacts are incorporated as a temporal correction by augmenting historical weather profiles (e.g., temperature and extreme events) with long-term trends. This enables a more accurate representation of their effects on system demand, transmission capacity, generator forced outages, and RES characteristics.
\end{enumerate}

The rest of this paper is organized as follows: Section II presents the proposed TRACED method; Section III conducts comprehensive case studies to evaluate the effectiveness of the proposed method in CC allocation to generation resources of varying size, class, and location; Section IV provides a conclusive discussion and potential future directions.

\section{The Proposed TRACED Method}
The ELCC calculation model is formulated as a transmission-constrained UC problem, with the objective of minimizing the system operational costs, while adhering to prevailing operational constraints. The reliability component is integrated into the problem by accounting for load shedding penalization. System load and weather conditions are modeled based on historical data while adjusted for the capacity market year to be solved, considering their yearly evolving trends.

\vspace{-10pt}
\subsection{Problem Formulation}

\subsubsection{Objective Function}
The objective function (\ref{eq:cost_obj}) is to minimize power production, start-up, and shut-down costs of thermal generators, power production and curtailment costs of renewables, battery charge costs, and load shedding costs. 

\vspace{-10pt}

\begin{equation}
\begin{aligned}
\min ~ 
& \sum \nolimits_{t \in T} \Bigl[
      \sum \nolimits_{i \in \mathcal{I}}
      \bigl( G_{i,t} \hspace{-1mm}\cdot \hspace{-1mm}C_i \hspace{-1mm}+ \hspace{-1mm}su_{i,t} \hspace{-1mm}\cdot\hspace{-1mm} SUC_i \hspace{-1mm}+ \hspace{-1mm}sd_{i,t} \hspace{-1mm}\cdot\hspace{-1mm} SDC_i \bigr) \\
& \quad \quad  ~
    + \sum \nolimits_{g \in \mathcal{\{S,W\}}}
      \bigl( G_{g,t} \cdot C_g + curt_{g,t}\cdot C^{\text{curt}}_{g} \bigr) \\
& \quad \quad  ~ 
    + \sum \nolimits_{b\in \mathcal{B}}
    \bigl(CH_{b,t} \cdot C_{b}\bigr)
    + LS_t \cdot VOLL
  \Bigr]
\label{eq:cost_obj}
\end{aligned}
\end{equation}

The production costs of thermal generators are based on their respective price curves, commonly formulated as piecewise linear functions \cite{S. Wang 2022}. Costs associated with operating solar, wind, and battery resources are assigned based on their respective levelized cost of energy (LCOE). RESs are treated as must-take resources, such as in CAISO \cite{L. Bird 2014}, by allowing curtailment but at penalties higher than the generation cost of regular resources but lower than the load shedding cost. 

\subsubsection{Power Balance Constraint}
Hourly system balance is enforced to ensure that the summed thermal, renewable, and battery generation equals the demand minus load shedding. 
\begin{equation}
\begin{split}
&\sum\nolimits_{n \in \mathcal{N}} (D_{n,t}-LS_{n,t}) =
\sum\nolimits_{g \in \{\mathcal{I},\mathcal{S},\mathcal{W}\}} G_{g,t} \\
&  \quad   \quad  \quad  \quad + \sum\nolimits_{b \in \mathcal{B}} \bigl( DIS_{b,t} - CH_{b,t} \bigr),
\quad \forall t \in \mathcal{T}
\end{split}
\label{eq:power_balance}
\end{equation}

\subsubsection{Transmission Constraints}
DC power flow is adopted to model the transmission network constraints. The power transmission distribution factor \cite{PTDF} is used to calculate the total real power transfer along each line and is limited to its maximum transfer capacity (\ref{eq:trans_limits}).
\begin{equation}
\begin{aligned}
\Biggl| 
& \sum \nolimits _{{g \in \{\mathcal{I},\mathcal{S},\mathcal{W}\}}} PTDF_{n(g),l} \, G_{g,t} \hspace{-1mm} + \hspace{-1mm}\sum \nolimits _{b \in \mathcal{B}} PTDF_{n(b),l} \,\cdot\\
& (DIS_{b,t} - CH_{b,t}) 
\hspace{-1mm} - \hspace{-1mm}\sum \nolimits _{n \in \mathcal{N}} PTDF_{n,l} \, \cdot (D_{n,t} \hspace{-1mm}-\hspace{-1mm} LS_{n,t}) 
\Biggr| \\
& \le C^{\max}_l \, \cdot AAR_{l,\tau_{\text{air}}}, \quad \forall l \in \mathcal{L}, \ \forall t \in \mathcal{T}
\label{eq:trans_limits}
\end{aligned}
\end{equation}

Transmission line capacities could be derated based on ambient air temperature $AAR_{l,\tau_{\text{air}}}$, referred to as ambient-adjusted rating (AAR). The use of hourly AAR is required by RTOs for studies \textless 10 days away (e.g., the day-ahead market) according to FERC Order 881. While the FERC Order allows for seasonal AAR ratings for long-term applications (e.g., the capacity market), it is justifiable to use the hourly AAR in this work as we sample and solve multiple UC problems with varying hourly-weather conditions. Several RTOs, including PJM and ERCOT, have implemented hourly AARs as a step-function for ambient temperature ranges \cite{FERC AAR Report}, and a similar process is followed by applying an AAR derating coefficient. 

\subsubsection{Thermal Generator Modeling}
Thermal generators must operate within their upper and lower capacity limits (\ref{eq:therm_gen_limits}), which are related to a temperature-dependent forced outage rate (FOR) $FOR_{i,\tau_{\text{air}}}$. In this paper, FOR of thermal generators is approximated via continuous FOR functions derived by applying a  4\textsuperscript{th} order polynomial regression to the discrete FOR data \cite{S. Murphy 2020}, and is used to derate the maximum thermal generator output, which is found not to significantly impact the CCs of renewables in the system \cite{S. Wang 2022}. State transitions are implemented to ensure that thermal generators' on/off variables are aligned with their coinciding start-up and shut-down variables (\ref{eq:therm_su_sd_constraint}). Additionally, minimum on and off times are guaranteed via (\ref{eq:therm_min_on_time}) and (\ref{eq:therm_min_off_time}), respectively. 

\vspace{-10pt}

\begin{subequations}
\begin{flalign}
&G^{\min}_i \, \cdot u_{i,t}
\le G_{i,t}
\le G^{\max}_i \,\cdot u_{i,t} \cdot
\bigl( 1 - FOR_{i,\tau_{\text{air}}} \bigr), \nonumber \\
&\qquad\qquad\qquad\qquad\qquad\qquad \ \forall i \in \mathcal{I},\ \forall t \in \mathcal{T} 
\label{eq:therm_gen_limits}\\
&u_{i,t} - u_{i,t-1} = su_{i,t} - sd_{i,t}, \ \forall i \in \mathcal{I}, \ \forall t \in [2,\mathcal{T}] 
\label{eq:therm_su_sd_constraint}\\
&\sum \nolimits _{\tau = t-UT_i+1}^{t} su_{i,t} \le u_{i,t},
 ~~\ \forall i \in \mathcal{I}, \ \forall t \ge UT_i
\label{eq:therm_min_on_time} \\
&\sum\nolimits_{\tau = t-DT_i+1}^{t} sd_{i,t} \le 1 \hspace{-1mm}- \hspace{-1mm}u_{i,t}, 
\hspace{-0.4mm} \forall i \in \mathcal{I}, \ \forall t \ge DT_i
\label{eq:therm_min_off_time}
\end{flalign}
\end{subequations}

\subsubsection{Renewable Generation}
This paper considers two major types of RESs, including solar photovoltaics (PVs) and wind farms, whose outputs are highly dependent on weather conditions. Specifically, PVs are impacted by temperature and solar irradiance, and wind farm performance relies on wind speed. 

PV cell temperature is modeled as (\ref{eq:solar_temp}) where \textit{NOCT} is the nominal operating cell temperature, and the maximum PV output under a specific weather condition is modeled as (\ref{eq:solar_gmax}). 

\vspace{-10pt}

\begin{subequations}
\begin{flalign}
&\tau_{s,t} = \tau_{\text{air},t} 
\hspace{-1mm}+ \hspace{-1mm}\frac{(NOCT_s - 20)}{800} \, \cdot Ins_{s,t}, 
\forall s \in \mathcal{S}, \hspace{-1mm}~ \forall t \in \mathcal{T}
\label{eq:solar_temp}\\
&G^{\max}_{s,t} 
= \gamma_s \, \frac{Ins_{s,t}}{1000} 
\Bigl[ 1 \hspace{-1mm}- \hspace{-1mm}\alpha_s (\tau_{s,t} \hspace{-1mm}- \hspace{-1mm}25) \Bigr] 
\eta_s,
\forall s \in \mathcal{S}, \hspace{-1mm}~\forall t \in \mathcal{T}
\label{eq:solar_gmax}
\end{flalign}
\end{subequations}

The maximum power output of a wind turbine is modeled as a piecewise function dependent on wind velocity \textit{v} (\ref{eq:wind_gmax}), where c\textsubscript{3} through c\textsubscript{0} of each turbine are defined according to a cubic regression fitted to the turbine’s power curve data in the corresponding operation region, and $v^{ci}_{w}$, $v^{r}_{w}$, $v^{co}_{w}$ are the cut-in, rated, and cut-out wind speeds of wind turbine $w$.

\vspace{-10pt}

\begin{equation}
\begin{aligned}
G^{\max}_{w,t}(v) &= \eta_w
\begin{cases}
0, & [0,v^{ci}_{w}) \\
c_3 v^3 + c_2 v^2 + c_1 v + c_0, &  [v^{ci}_{w}, v^{r}_{w}) \\
\gamma_w, & [v^{r}_{w}, v^{co}_{w}) \\
0, & [v^{co}_{w}, \infty)
\end{cases} \\
& \qquad\qquad\qquad\qquad\qquad~~ \forall w \in \mathcal{W}, \ \forall t \in \mathcal{T}
\end{aligned}
\label{eq:wind_gmax}
\end{equation}

To account for the situation where transmission congestion prevents full power delivery from RESs, curtailment is allowed as in (\ref{eq:solar_curt}) and (\ref{eq:wind_curt}). Additionally, should a hurricane or tropical storm impact the location of a wind farm, the turbines are forced to stall with no power output as described by binary variable \textit{hurr} =1.
\begin{subequations}
\begin{flalign}
&curt_{s,t} = G^{\max}_{s,t} - G_{s,t} \ge 0, \quad \forall s \in \mathcal{S}, \ \forall t \in \mathcal{T}
\label{eq:solar_curt} \\
&curt_{w,t} = G^{\max}_{w,t} \cdot (1 - hurr_{w,t}) - G_{w,t} \ge 0, \nonumber\\
& \qquad\qquad\qquad\qquad\qquad\qquad ~~\forall w \in \mathcal{W}, \ \forall t \in \mathcal{T}
\label{eq:wind_curt}
\end{flalign}
\end{subequations}

\subsubsection{Energy Storage Constraints}
A battery energy storage system (BESS) is constrained by its min and max capacities (\ref{eq:soc_limits}), charging and discharging limits (\ref{eq:storage_charge})-(\ref{eq:storage_discharge}), state transition (\ref{eq:soc_dynamic}), and possible charging/discharging/idle operation status in each time step (\ref{eq:charge_discharge_exclusive}). 
\begin{subequations}
\begin{flalign}
&SOC^{\min}_b \le SOC_{b,t} \le SOC^{\max}_b, 
~\forall b \in \mathcal{B}, \ \forall t \in \mathcal{T}
\label{eq:soc_limits} \\
&0 \le CH_{b,t} \le CH^{\max}_b \, \cdot ch_{b,t}, 
\quad ~~\forall b \in \mathcal{B}, \ \forall t \in \mathcal{T}
\label{eq:storage_charge} \\
&0 \le DIS_{b,t} \le DIS^{\max}_b \, \cdot dis_{b,t}, 
~~ \forall b \in \mathcal{B}, \ \forall t \in \mathcal{T}
\label{eq:storage_discharge}\\
&SOC_{b,t} \hspace{-1mm} \cdot\hspace{-1mm}  E^{max}_b = SOC_{b,t-1} \hspace{-1mm} \cdot \hspace{-1mm} E^{max}_b
+ CH_{b,t} \, \hspace{-1mm} \cdot \hspace{-0.5mm} \eta_{b}^{\text{char}} \nonumber \\
&\qquad\qquad- {DIS_{b,t}}/{\eta_{b}^{\text{dis}}},  ~~~~\qquad \forall b \in \mathcal{B}, \ \forall t \in [2, \mathcal{T}]
\label{eq:soc_dynamic}\\
&ch_{b,t} + dis_{b,t} \le 1, 
~~~~~~~~~~~~~~~\quad \forall b \in \mathcal{B}, \ \forall t \in \mathcal{T}
\label{eq:charge_discharge_exclusive}
\end{flalign}
\end{subequations}

\subsubsection{Reliability Constraints}
To integrate reliability constraints into the UC framework, load shedding due to insufficient generation and/or transmission capacities is allowed and limited to no more than the total system demand (\ref{eq:sys_load_shedding}). Additionally, load shedding is distributed to each bus in the system according to its associated load weight ($\omega_n$) \ref{eq:bus_load_shedding}. 
\begin{subequations}
\begin{flalign}
&0 \le LS^{sys}_t \le ls_t \cdot \sum \nolimits _{n \in \mathcal{N}} D_{n,t}, 
\quad \forall t \in \mathcal{T}
\label{eq:sys_load_shedding} \\
&LS_{n,t} = \omega_n \, \cdot LS^{sys}_t, 
\quad\quad\quad\quad\quad~~ \forall n \in \mathcal{N}, \ \forall t \in \mathcal{T}
\label{eq:bus_load_shedding}
\end{flalign}
\end{subequations}

\subsubsection{UC Model with Reduced-Complexity}
To solve TRACED efficiently, several typical short-term UC constraints are simplified. Thermal generator ramping limits are removed, which is common in chronological Monte Carlo sampling \cite{EPRI 2024}; Hot and cold reserve requirements, which primarily buffer generator outages \cite{Hedman 2010}, are omitted because these contingencies are already captured in the FOR formulation.

Long-term system characteristics, such as load profiles and RES outputs, primarily exhibit monthly and seasonal patterns, motivating TRACED to be solved monthly, which is also consistent with NYISO’s monthly capacity market auctions. However, solving the hourly UC problem over a month is still computationally intractable. To balance complexity and temporal fidelity, a rolling-time horizon is implemented, well established in many UC studies \cite{S. Riaz 2019, G. Erichsen 2019}. A 1-week horizon with a 1-day overlap is adopted, sufficient to respect the minimum on/off time requirements of all generators.

\subsubsection{Weather and Load Modeling}
The common industry practice for ELCC evaluation relies on historical weather data to predict future load and generation, while considering demand variations due to major system developments. However, historical weather data may not reflect evolving climate trends, which affect both load and generation, particularly weather-sensitive RESs. This work incorporates such evolving weather trends to adjust system load, generator outputs, and forced outages, enabling more accurate system characterization and ELCC estimation in future operational environments. 

Hourly weather data of multiple historical years establish the baseline for temperature, solar irradiance, wind speed, and storm occurrence profiles, alongside corresponding system loads. To capture long-term temperature evolution, monthly average temperatures are regressed over time, with slope $\beta_{\tau,m}$ describing the warming/cooling trend of month \textit{m}. For an evaluation year \textit{y\textsuperscript{*}} and month \textit{m}, a sampled historical year \textit{y} is adjusted via (\ref{eq:adjusted_air_temp}), and the resulting hourly temperature profile is used to modify generator FORs, PV output, and line AARs as detailed below.
\begin{equation}
\tau^{\text{adj}}_{\text{air},t} 
= \tau^{\text{samp}}_{\text{air},t} 
+ \beta_{\tau,m} \, \cdot (y^* - y),
\quad \forall t \in \mathcal{T}
\label{eq:adjusted_air_temp}
\end{equation}

A piecewise linear regression captures the sensitivities of hourly system load to temperature for different temperature ranges (e.g., heating- vs. cooling-driven demand). This model is used to project system load under future temperature scenarios. The trend-adjusted temperature is mapped through the load–temperature regression to adjust the sampled demand in (\ref{eq:adjusted_demand}), where $D_{LR}(\cdot)$ denotes the piecewise regression and $D^{\text{samp}}$ is the projected (but not weather-adjusted) system demand. This approach preserves daily, weekly, and seasonal load variability while incorporating long-term trends.

\vspace{-10pt}

\begin{equation}
D^{\text{adj}}_{t} 
= D^{\text{samp}}_{t} 
+ \Bigl[ D_{LR}(\tau^{\text{adj}}_{\text{air},t}) - D_{LR}(\tau^{\text{samp}}_{\text{air}, t}) \Bigr], 
\quad \forall t \in \mathcal{T}
\label{eq:adjusted_demand}
\end{equation}

In addition to temperature adjustments, the weather model also captures the sequential impacts of hurricanes on RES availability, considering their long-term occurrence trends, via a probabilistic model. Historical records are used to estimate frequency and duration of storm events, and a linear regression on the annual storm frequency yields slope $\beta_{\text{hurr}}$ to represent its long-term trends. Monthly storm counts affecting the studied region $H_m$ are adjusted via $\beta_{\text{hurr}}$ and normalized by the number of historical years $N_{\text{yrs}}$ and hours $HRS_m$ in evaluation month \textit{m} to obtain the hourly storm probability in \eqref{eq:hurricane_prob}.
\vspace{-1pt}
\begin{equation}
\text{Prob}_{\text{hurr},t} 
= \frac{H_m + (y^* - y) \, \cdot \beta_{\text{hurr}}}{N_{\text{yrs}} \,\cdot HRS_m}, 
\quad \forall t \in \mathcal{T}
\label{eq:hurricane_prob}
\end{equation}

A Bernoulli trial is conducted using \eqref{eq:hurricane_prob} to determine whether a storm event occurs each hour. When a storm is triggered at time step $t_{hit}$, wind farms are forced into the stall mode for a duration determined by a normal distribution (with mean $\mu_{m}$ and standard deviation $\sigma_{m}$ for month $m$) of historical outage lengths, with an additional buffer period $Buff$ applied to capture pre-event preparations and post-event recovery (\ref{eq:hurricane_duration}). 
\begin{equation}
Durr_{\text{hurr}} = \text{norm}(\mu_m, \sigma_m) + Buff,
\label{eq:hurricane_duration}
\end{equation}

\vspace{-5pt}

This outage window is centered on $t_{hit}$, and any outage periods extending beyond the modeled horizon [1,…,T] are truncated accordingly as in (\ref{eq:hurricane_window}).

\vspace{-10pt}

\begin{equation}
\begin{aligned}
& hurr_{w,t} = 1,
\forall w \in \mathcal{W}, \\
& ~~~~\forall t \in \Bigl[ t_{\text{hit}} - {Durr_{\text{hurr}}}/{2}, \ 
t_{\text{hit}} + {Durr_{\text{hurr}}}/{2} \Bigr] \cap [1, T]
\end{aligned}
\label{eq:hurricane_window}
\end{equation}

\vspace{-30pt}
\subsection{TRACED Workflow}
The workflow of the proposed TRACED method is described in Algorithm \ref{alg:traced_workflow}, including the following stages: 

\begin{algorithm}[!b]
\caption{TRACED Workflow for CC Evaluation}
\label{alg:traced_workflow}
\begin{algorithmic}[1]

\STATE \textbf{Stage 1. Define bulk energy system environment:} 
\STATE \quad Characterize thermal generators and transmission \\ \quad \quad constraints;
\STATE \quad Identify RESs and BESSs for CC evaluation.
\STATE \textbf{Stage 2. Historical sampling and climate adjustment:}
\STATE \quad Sample $Y$ historical years of load and weather data;
\STATE \quad Apply temperature adjustments (\ref{eq:adjusted_air_temp}) and projected \\ \quad \quad extreme weather trends (\ref{eq:hurricane_prob}) - (\ref{eq:hurricane_window});
\STATE \quad Apply temperature-correlated adjustments to load  (\ref{eq:adjusted_demand}), \\ \quad \quad transmission AARs (\ref{eq:trans_limits}), and PV outputs (\ref{eq:solar_temp}). 

\STATE \textbf{Stage 3. Compute load adjustment:}
\FOR{each system type (base, FI, LI, portfolio)}
    \STATE Initialize $LA$, $LA_{\max} = \infty$, and $LA_{\min} = -\infty$;
    \STATE Add RESs/BESSs for this system;

    \WHILE{$LOLE_{\text{LA}} \neq LOLE_{\text{tar}}$ \OR $LA_{\max}-LA_{\min} > \epsilon_{\text{LA}}$}
        \STATE Solve hourly transmission-constrained UC across all \\  \quad sub-periods of $Y$ sampled years (e.g., rolling \\  \quad weekly solves, carry over initial/final conditions);
        \STATE Compute $LOLE_{\text{LA}} = \frac{1}{Y} \sum_{y=1}^{Y} LOLE_y$;
        
        \IF{$LOLE_{\text{LA}} > LOLE_{\text{tar}}$ \AND $LA < LA_{\max}$}
            \STATE $LA_{\max} \gets LA$;
        \ELSIF{$LOLE_{\text{LA}} < LOLE_{\text{tar}}$ \AND $LA > LA_{\min}$}
            \STATE $LA_{\min} \gets LA$;
        \ENDIF
        
        \IF{$LA_{\min} = -\infty$ \OR $LA_{\max} = \infty$}
            \STATE Adjust $LA$ up or down until both bounds are finite;
        \ELSIF{$LA_{\min} \neq -\infty$ \AND $LA_{\max} \neq \infty$}
            \STATE $LA \gets (LA_{\min} + LA_{\max}) / 2$ ;
        \ENDIF
    \ENDWHILE
\ENDFOR

\STATE \textbf{Stage 4. Compute CCs via Eqs. (\ref{eq:PORT_def}) - (\ref{eq:delta_elcc})}
\end{algorithmic}
\end{algorithm}

\textbf{Stage 1. System definition:} The bulk system operation environment is defined, including the month \textit{m} and year \textit{y*} for CC evaluation. Thermal generators are characterized, including capacities, minimum on/off times, locations, and FORs. The transmission system and associated line capacities are defined. RESs and BESSs for CC evaluation are identified.

\textbf{Stage 2. Historical sampling and climate adjustment: } Weather and operational conditions for \textit{Y} historical years are sampled to obtain system nodal load and weather data (temperature, solar irradiance, wind speed, and extreme weather events). Future climate trends are incorporated to adjust the system demand, line AARs, and generator outputs and FORs of the CC evaluation year.

\textbf{Stage 3. Load adjustment: } For each system configuration (base, portfolio, FI, and LI systems), the load adjustment (LA) to satisfy the target LOLE is determined. A search-and-bound procedure is used to identify upper and lower bounds for LA, denoted $LA_{max}$ and $LA_{min}$, such that they produce a LOLE above and below the target reliability level. The interval is iteratively narrowed until the target $LOLE_{tar}$ is achieved within a specified tolerance $\epsilon_{LA}$. This procedure is performed across all \textit{Y} sampled profiles, and the average LA is used for subsequent calculations.

\textbf{Stage 4. Generator CC evaluation:} Using the LA that satisfies the LOLE criterion for: (i) the base system with no generators of interest, $LA_{Base}$; (ii) the portfolio system with all generators of interest, $LA_{Port}$; (iii) all FI systems with only generator $j$, $LA_{FI_j}$; and (iv) all LI systems with all generators except generator $j$, $LA_{LI_j}$. The CC of generator $j$ is computed using the Delta method (\ref{eq:PORT_def}) - (\ref{eq:delta_elcc}). 
\begin{subequations}
\begin{flalign}
&PORT = LA_{Port}-LA_{Base},
\label{eq:PORT_def} \\
&FI_{j} = LA_{FI_j}-LA_{Base}, \quad \forall j \in \mathcal{J}
\label{eq:FI_def} \\
&LI_{j} = LA_{Port}-LA_{LI_j}, \, \, \quad \forall j \in \mathcal{J}
\label{eq:LI_def} \\
&PIE = PORT - \sum \nolimits _{j \in \mathcal{J}} LI_j,
\label{eq:pie_def}\\
&IIE_j = FI_j - LI_j,
\, \, \quad\quad\quad \forall j \in \mathcal{J}
\label{eq:iie_def}\\
&\delta = {PIE}/{\sum\nolimits_{j \in \mathcal{J}} IIE_j},
\label{eq:delta_def}\\
&\text{ELCC}_j = LI_j + \delta \, IIE_j,
\, \, \quad \forall j \in \mathcal{J}
\label{eq:delta_elcc}
\end{flalign}
\end{subequations}

\vspace{-10pt}
\section{Case Studies}
To evaluate the performance of the proposed TRACED method, comprehensive case studies are conducted to compare the CCs of RESs and BESSs across months in different seasons using the last-in marginal ELCC (i.e., industry practice) and TRACED. Sensitivity analyses are also performed to better understand the impacts of transmission congestion and forward-looking weather modeling on CC evaluation. 

\vspace{-10pt}
\subsection{Simulation Setup}
\vspace{-3pt}
The simulation is built upon the IEEE 118-bus system \cite{Alroomi 2015}, which includes thermal generator characteristics (e.g., maximum/minimum outputs, minimum on/off times, pricing curves, start-up and shut-down costs, and their locations in the system) and transmission topology. Costs associated with RESs/BESSs are assigned based on their respective LCOEs \cite{AEO 2025}, and the curtailment cost of RESs is assigned 10 times its LCOE. Cost of load shedding is \$10,000/MWh, per MISO \cite{MISO VOLL}. These system costs are summarized in \cite{Appendix}.

\begin{figure}[h]
\centering
\vspace{-10pt}
\includegraphics[width=2in]{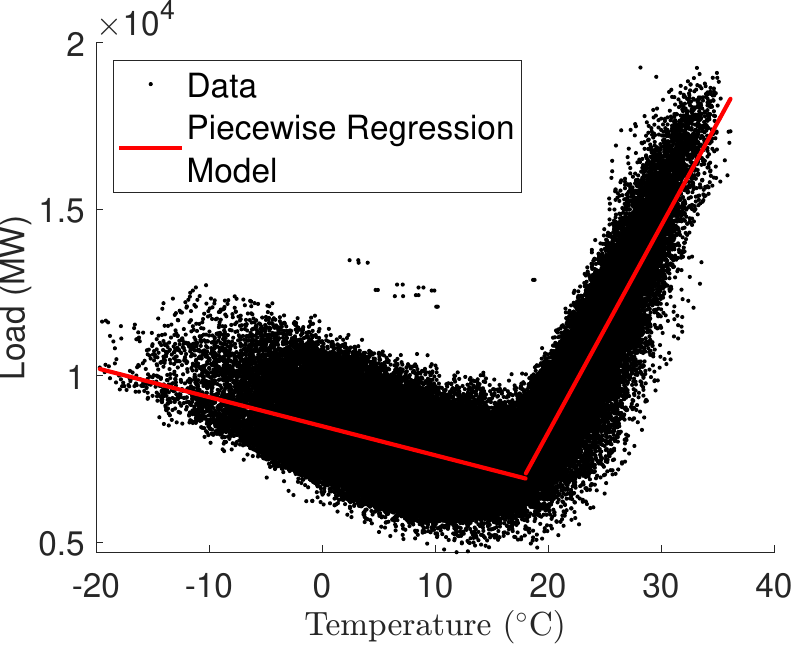}
\vspace{-10pt}
\caption{Load vs temperature piecewise regression modeling.}
\vspace{-12pt} 
\label{Load vs Temperature}
\end{figure}

Load profile of the IEEE 118-bus system is modified based on New Jersey's energy consumption trajectory (1993-2023) \cite{PJM Hourly Load}. All transmission lines’ power transfer limits are adjusted based on ambient temperature derating \cite{Allowable Ampacity Tables}. Historical weather data from 1993 to 2023 is collected: solar irradiance and air temperature data of New Jersey were collected from Open-Meteo \cite{Open-Meteo}; wind speed data were collected from the NREL Wind Resource Database \cite{Wind Resource Database} at the location of the Ocean Wind 1 farm and the Jersey-Atlantic City Wind Farm for offshore and onshore wind resources, respectively; hurricane occurrence/duration data impacting New Jersey were collected from the NOAA Historical Hurricane Tracks database \cite{Historical Hurricane Tracks}. Year-to-year trends in average monthly temperature and yearly hurricane occurrence were extrapolated from this data to apply the future-looking weather adjustments \cite{Appendix}. The piecewise regression model for hourly load vs temperature correlation is developed using their historical data as shown in Figure \ref{Load vs Temperature}, and then the future-weather adjustment (\ref{eq:adjusted_demand}) is applied. 

With the above data of 31 weather years and 403 weekday-shifted load profiles, 100 monthly weather and load profiles were sampled via Monte Carlo simulations. Since TRACED is solved hourly over one-month time spans, the standard LOLH of 2.4 hours/year is converted to 0.2 hours/month. Such seasonal or monthly LOLH has been implemented in ISO’s reliability studies \cite{MISO 2025-2026 Study Report}. Although these studies typically assign different seasonal values, with the most at-risk periods (typically summer) having a higher LOLH, this study assigns each month the same value in considering that at-risk seasons/months may shift with the transition of energy systems. Simulations were performed in MATLAB R2025a and Gurobi 12.0.3 on a Linux workstation equipped with an AMD Ryzen Threadripper Pro 7985WX processor and 528 GB of RAM. 

\vspace{-5pt}
\subsection{Case Study Designs}
\vspace{-3pt}
Four cases with different portfolios of RESs and BESSs are built to illustrate TRACED's performance under various system environments: 
(i) \textbf{Case 1}, a portfolio of three wind farms (1 offshore and 2 onshore) totaling 3,700MW of installed capacity (30.1\% of system installed capacity);  
(ii) \textbf{Case 2}, a portfolio of three solar farms totaling 3,700MW; 
(iii) \textbf{Case 3}, combining the portfolios of wind and solar farms in Cases 1 and 2, totaling 7,400 MW of installed RES capacity (46.3\% of system installed capacity);
and (iv) \textbf{Case 4}, built on Case 2 by adding 3 BESSs totaling 1850MW, for a combined portfolio capacity of 5550MW (39.0\% of system installed capacity).

\begin{figure}[h]
\vspace{-15pt}
\centering
\includegraphics[width=3in]{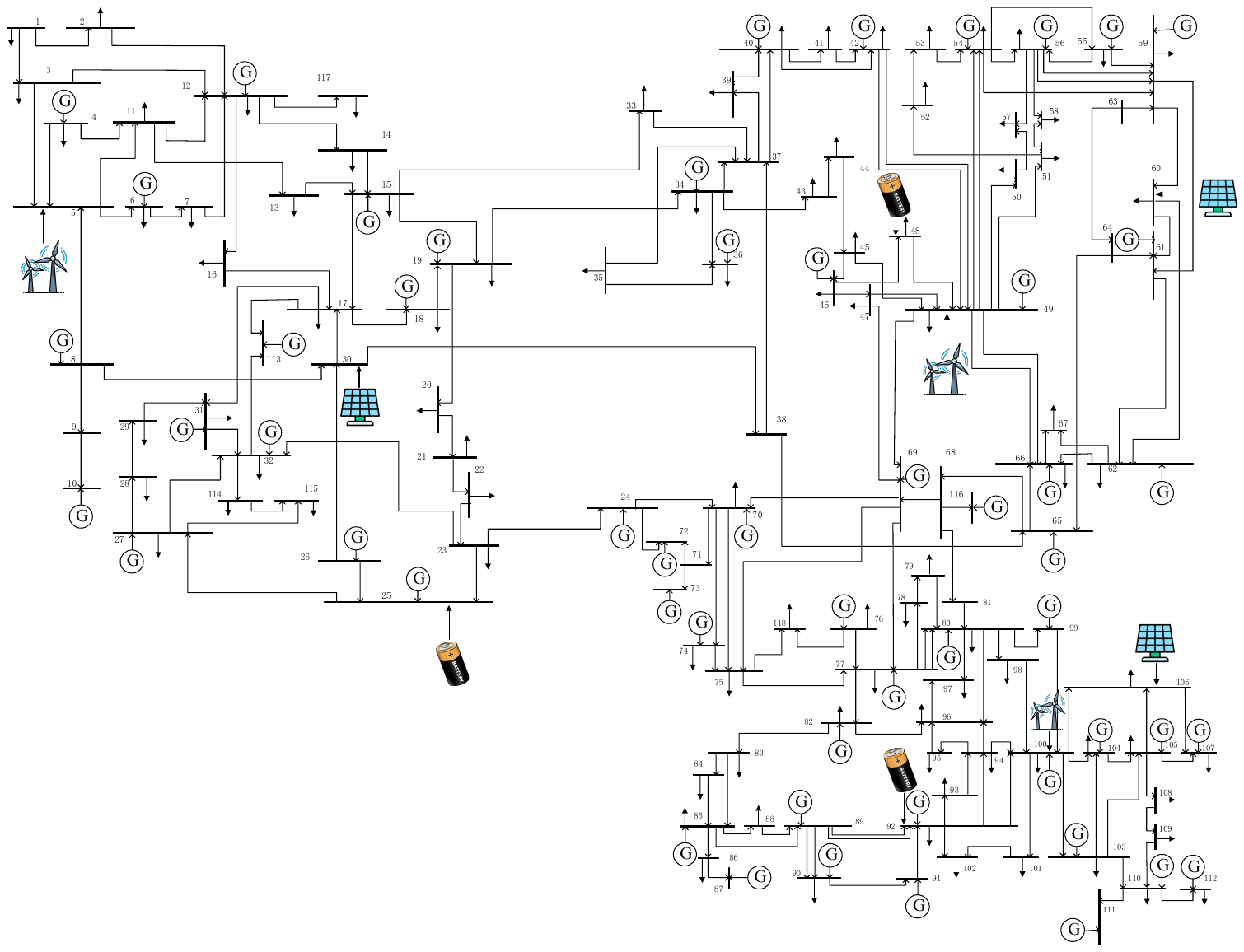}
\vspace{-20pt}
\caption{Modified IEEE 118-bus system. }
\label{Modified 118-bus system}
\end{figure}

Cases 1 and 2 are designed to evaluate how individual generators under the same resource class impact CCs of each other; Case 3 further assesses how environmentally sensitive RESs that tend to peak at different times (i.e., wind and irradiance) could impact their respective CCs; 
 Case 4 unveils how storage, as a complementary resource to RESs, can help improve overall system reliability and impact CCs of other RESs, which is most representative of what future energy systems will likely resemble. The CCs of RESs and BESSs in all cases are evaluated across three months (June, December, and April) for year 2030 to illustrate the impacts of seasonal variations in load and weather. Key parameters of these four cases are summarized in Table \ref{tab:case_params}, and the modified 118-bus system is illustrated in Figure \ref{Modified 118-bus system}.

\begin{table}[h]
\centering
\vspace{-10pt}
\caption{Key Parameters of Case Studies}
\vspace{-5pt}
\label{tab:case_params}
\begin{tabular}{l p{2.2cm} c c}
\toprule
\textbf{Case} & \textbf{RES/Battery \#} & \textbf{Rated Capacity} & \textbf{Bus Location} \\
\midrule
\multirow{2}{*}{Case 1} 
    & Wind 1/2 (onshore) & 500/1,200 MW & 5/49 \\
    & Wind 3 (offshore) & 2,000 MW & 100 \\
\midrule
\multirow{1}{*}{Case 2} 
    & Solar 1/2/3 & 500/1,200/2,000 MW & 30/60/106 \\
\midrule
Case 3 & \multicolumn{3}{l}{Case 1 and Case 2 Generators} \\
\midrule
\multirow{4}{*}{Case 4} 
    & Case 2 Generators &  &  \\
    & BESS 1 & 700 MWh / 350 MW & 25 \\
    & BESS 2 & 1000 MWh / 500 MW & 48 \\
    & BESS 3 & 2000 MWh / 1000 MW & 92 \\
\bottomrule
\end{tabular}
\vspace{-11pt}
\end{table}

Two sets of sensitivity analyses are further performed:

\noindent (1) \textit{Sensitivity test of transmission congestion}: For \textbf{Test 1.1}, Solar Farm 3 and Wind Farm 3, as the two RESs whose CCs are mostly influenced by transmission congestion, are selected for analysis, and the power transfer capacity of the lines connected to their buses (i.e., lines 167, 169, and 172 for Solar Farm 3; and lines 154, 155, 158-160, 163, 164, and 167 for Wind Farm 3) are upgraded to assess the impacts on their CCs; For \textbf{Test 1.2}, lines 128, 129, and 155, identified as the most congested branches of the system, as well as lines 37, 147, 153, and 169, as the next-level of congested branches of the system, are upgraded and the overall system reliability is reevaluated. While impractical, a full system upgrade test is also conducted to highlight the potential reliability improvements should the transmission system no longer be a bottleneck. 
These sensitivity analyses are performed on a portfolio of all 9 wind, solar, and BESSs using the TRACED method for the summer month of June, which is the most congested season in the current system setting.

\noindent (2) \textit{Sensitivity test of future weather trend modeling}: \textbf{Test 2.1} evaluates the impact on solar farms' CCs by adjusting the year-to-year temperature trend under Case 2 settings, and \textbf{Test 2.2} evaluates the impact on wind farms' CCs by varying year-to-year hurricane occurrence trends under Case 1 settings, as their outputs and FORs are mostly influenced by ambient temperature and hurricane occurrence, respectively.

\vspace{-8pt}
\subsection{Case Study Results}
\subsubsection{Case 1}
\begin{figure}[h]
\centering
\vspace{-15pt}
\includegraphics[width=2.25in]{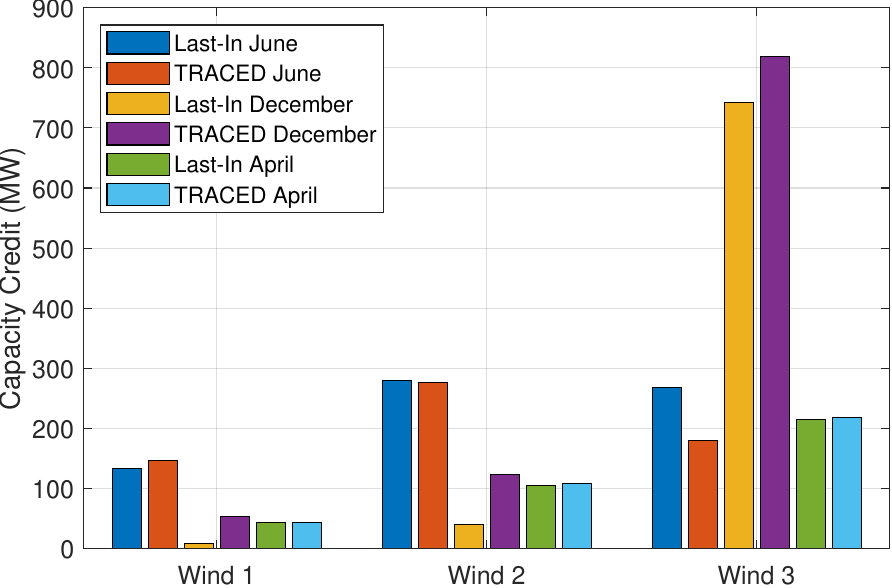}
\vspace{-10pt} 
\caption{CCs of wind generators in different months.}
\vspace{-5pt} 
\label{Case1_results}
\end{figure}

As shown in Figure \ref{Case1_results}, the CCs of wind farms vary significantly among different months, reflecting the seasonal variability of wind resources. In December, when wind speeds are at their peaks, the portfolio CC also hits the highest value of 995MW (26.9\% of nameplate capacity). However, this is not necessarily the case for individual generators. Instead, Wind Farms 1 and 2 exhibit low ELCCs despite high wind resources. The offshore Wind Farm 3 saturates the system with its supply, reducing Wind Farms 1 and 2's reliability contributions during the peak wind production hours due to the shifted net load. With the TRACED method (marked purple), the FI marginal ELCC of Wind Farm 3 is 881.8MW, while its LI marginal ELCC is 742.2MW, resulting in a positive IIE. Because of the positive PIE, those extra CCs get reallocated, increasing Wind Farm 3's CCs by 76.5MW (+10.3\%). Similar trends are observed for the other two generators, resulting in higher accreditations by TRACED than the LI marginal ELCC.  

In June, when wind resources are less, the portfolio CC is smaller than that in December at 603MW (16.2\%). Wind Farm 3, integrated by mostly congested lines, contributes less to the system reliability due to degraded transmission capacities, which become more impactful during the summer months with a high system load level and elevated ambient temperatures. As a result, the power produced by Wind Farm 3 cannot be fully delivered to the system loads, reducing its CCs. This benefits Wind Farms 1 and 2 because their locations within the system are not saturated with wind power, thus contributing more to meet the system peak demand and resulting in higher CCs than in December. Thus, wind farms, induced by transmission congestion, act as complementary, where their LI CCs are greater than their FI CCs, causing TRACED to accredit Wind Farms 2 and 3 less than the LI marginal ELCC.  

In April, both system load and wind resources are the lowest compared to June and December, resulting in the lowest portfolio CC at 370MW (10\%). From the individual generator's perspective, Wind Farm 3 does not have enough available wind resources to saturate the system. As such, Wind Farms 1 and 2 can supply their full output to the system, each at $\sim$10\% of their nameplate capacities.

In summary, TRACED accredits the portfolio 77.7MW less (-11.4\%) in June compared to the LI marginal ELCC, but 6.9MW (+1.9\%) and 206MW more (+26.1\%) in April and December, respectively, providing a net benefit.

\subsubsection{Case 2}
Solar farms' CCs also vary notably by month, reflecting the availability of solar resources and transmission capacities. In June, when solar irradiance is the highest, the portfolio CC peaks at 657MW (17.8\%). However, Solar Farm 3 has a similar CC as Solar Farm 2 despite a 67\% higher nameplate capacity, attributed to limited transmission capacities around Solar Farm 3. Because of the congestion surrounding Solar Farm 3, its power output cannot be fully delivered to other areas of the system. Adding other solar farms actually could improve Solar Farm 3's CCs by relieving stress in areas of the system where it could not supply on its own, resulting in a negative IIE. Because of the positive PIE, Solar Farm 3's CC under the TRACED method decreases by 11.4MW (-4.5\%) compared to the LI marginal ELCC.  

\begin{figure}[h]
\centering
\vspace{-5pt}
\includegraphics[width=2.25in]{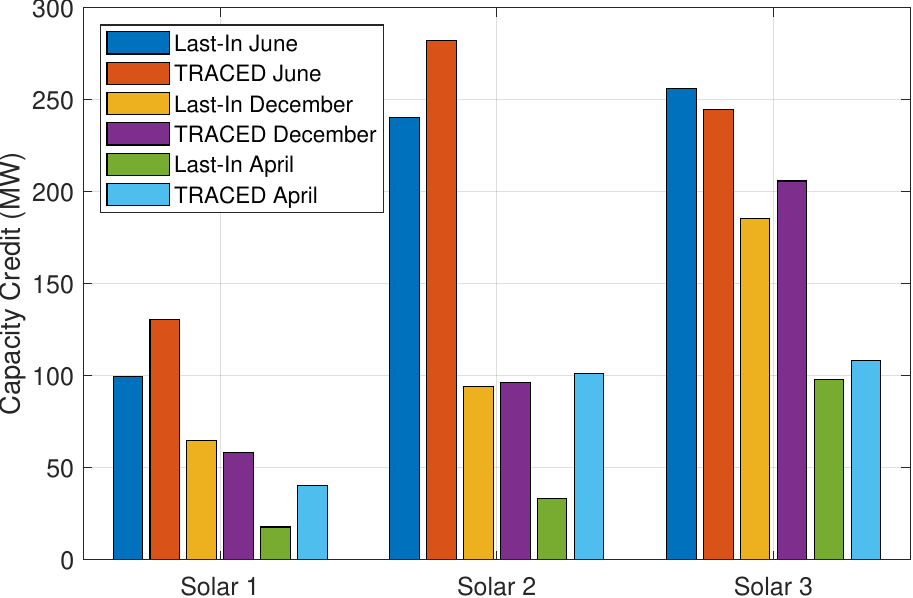}
\vspace{-10pt}
\caption{CCs of solar generators in different months.}
\vspace{-5pt} 
\label{Case2_results}
\end{figure}

In December, when solar irradiance is the lowest, the portfolio CC is less at 360MW (9.7\%), and individual generators’ CCs follow a more predictable pattern, scaled as the nameplate capacity increases, each at 8\% to 12\% of their nameplate capacity. This indicates that other factors (i.e., transmission congestion) do not play a dominant role in influencing the ELCC rating, which is expected as the ambient temperature derating improves the power transfer limit in colder months. 

In April, despite a higher resource availability than in December, the portfolio CC is the lowest at 249MW (6.7\%), mainly attributed to the lowest system load in shoulder seasons, as shown in Figure \ref{Load vs Temperature}. This justifies that resources provide minimum reliability benefits during off-peak demand periods.

In summary, TRACED accredits the portfolio 61.5MW more (+10.3\%) in June, 16.5MW more (+4.8\%) in December, and 100.5MW more (+67.7\%) in April compared to the LI marginal ELCC, providing a net benefit.

\subsubsection{Case 3}
Case 3 presents similar trends that the highest CCs are observed in winter for Wind units and in summer for Solar units. Nevertheless, notable differences are worth emphasizing: most CC ratings for individual generators decreased compared to Cases 1 and 2, except Solar Farm 2. The combined wind and solar portfolios' CC ratings from Cases 1 and 2 are 1,259.7MW, 1,355.4MW, and 619.2MW in June, December, and April, respectively, while are only 1,065.5MW, 1,090.9MW, and 544.1MW in Case 3. This CC reduction shows that reduced reliability contributions are obtained from additional resources added to the system. For example, Wind Farm 3 is connected to bus 100 and Solar Farm 3 to bus 106, only 1 branch away. Wind production in December saturates that area of the system, and Solar Farm 3 thus cannot contribute as much due to the congestion of branches 167, 169, and 172, decreasing its CC. Similarly, when Solar Farm 3 produces a high output to saturate the system in June, the reliability contribution of Wind Farm 3 is reduced. This highlights that the locations of RESs can significantly impact their CCs and nearby RESs should be carefully studied when identifying and assessing candidate RES installation sites.

\begin{figure}[h]
\centering
\vspace{-5pt}
\includegraphics[width=2.5in]{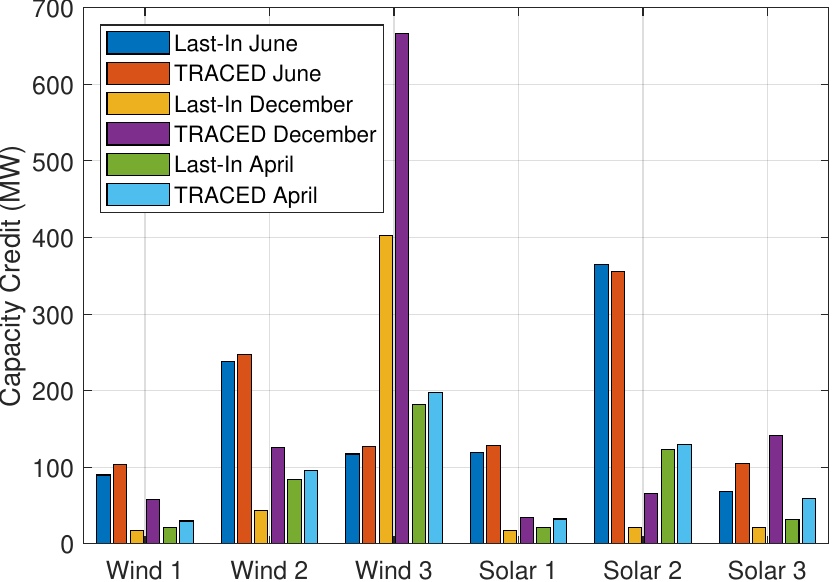}
\vspace{-5pt}
\caption{CCs of wind and solar generators in different months.}
\label{Case3_results}
\end{figure}

\vspace{-4pt}
In summary, TRACED accredits the entire portfolio with an additional 67.5 MW (+6.7\%) in June, 567.4 MW (+108\%) in December, and 81.2 MW (+17.5\%) in April relative to the LI marginal ELCC. From the perspective of individual generators, Wind Farm 3 shows the largest CC improvement in December under TRACED compared to the LI marginal ELCC. While its FI marginal CC remains unchanged, additional generators reduce its LI marginal CC to 402.3 MW. This results in the highest IIE, and consequently a greater allocation of the PIE, yielding a 264.4 MW (+65.7\%) increase in accredited capacity.

\subsubsection{Case 4}
Case 4 results highlight the reliability benefits of complementary technologies, such as BESS, for intermittent RES. In particular, in December and April, it is observed that when BESSs are added, the combined CCs of the solar farms increased by 269MW (+74.7\%) and 99.9MW (+40.1\%), respectively, compared to Case 2 with TRACED. This is because BESSs can reshape the load profiles, allowing excess generations to be dispatched at new net load peaks, improving the CCs of solar farms. Nevertheless, such complementary characteristics are not observed in June. The combined CCs of solar farms decreased by 130.5MW (-24.5\%) compared to Case 2, because the system is congested by the high demand and the reduced line capacities in higher ambient temperature, restricting the ability of BESSs to store excess PV generation and supply loads during those shifted net load peaks. This case demonstrates that the complementary characteristics are beneficial to resource reliability contributions, but can be restricted by transmission capacities, emphasizing the importance of integrating transmission constraints into CC evaluation. 

\begin{figure}[h]
\vspace{-5pt}
\centering
\includegraphics[width=2.5in]{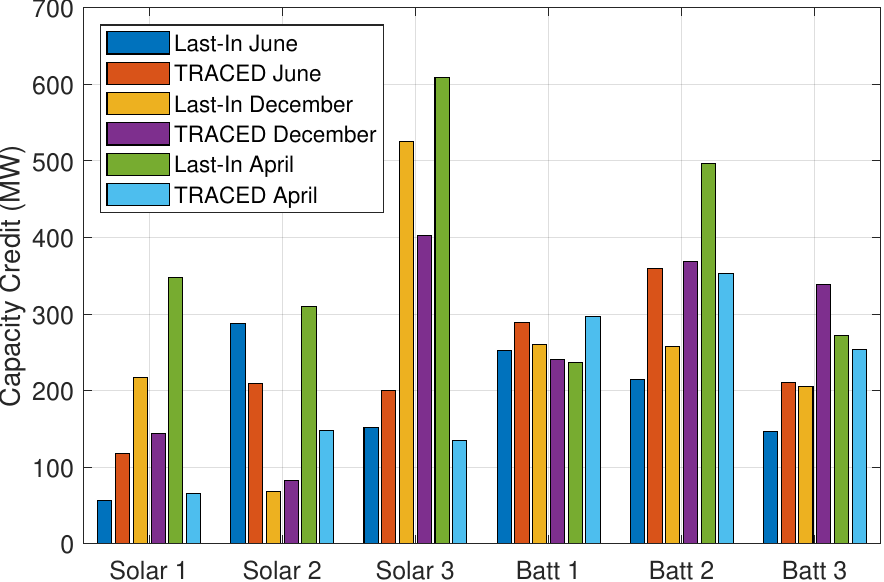}
\vspace{-8pt}
\caption{CCs of solar and BESS generators in different months.}
\vspace{-5pt}
\label{Case4_results}
\end{figure}

Due to these complementary characteristics observed in December and April, solar farms with BESSs add more reliability benefits than they would alone, resulting in LI CCs higher than FI CCs. Consequently, TRACED accredits the solar farms less than the LI ELCC. This occurs because the LI ELCC measures the improvement in reliability after adding the solar resource to a system that already contains BESSs, thus attributing the complementary reliability gains enabled by solar–storage interactions entirely to the solar resource evaluated last. When this is done across all solar farms, these complementary reliability gains are double-counted, where the sum of LI CCs exceeds the portfolio CC. TRACED avoids this double-counting effect by allocating CCs leveraging the system-wide reliability contribution, resulting in lower but portfolio-consistent CCs for the solar farms, reducing the likelihood of resource under-procurement. TRACED accredits solar farms 30.6 MW more (+6.2\%) in June, 181.6 MW less (-22.4\%) in December, and 918.7 MW less (-72.5\%) in April. Comparisons for individual generators using LI marginal ELCC and TRACED are shown in Fig.  \ref{Case4_results}.

While Solar Farms experience complementary benefits with the BESS integration, BESS CCs do not necessarily benefit from Solar Farms, with LI CCs less than their FI CC. This is because solar farm curtailment is highly penalized but not for BESSs, thus prioritizing RES generation over BESS discharge and impacting their CCs. TRACED tends to accredit resources between their FI and LI CCs, resulting in higher CCs than the LI marginal ELCC. TRACED accredits the BESS combined CCs 246.7MW (+40.2\%) and 225.1MW (+31.1\%) more in June and December, respectively, and 100.2MW (-10.0\%) less in April. Comparisons for individual BESS CCs using LI marginal vs TRACED can be seen in Figure \ref{Case4_results}.

In summary, TRACED accredits the portfolio 277.3MW more (+25.0\%) in June, 43.5MW more (+2.8\%) in December, and 1018.9MW less (-44.9\%) in April compared to the LI marginal ELCC.

\subsubsection{Sensitivity Tests of Transmission Congestion Impact}
\begin{figure}[h]
\vspace{-5pt}
\centering
\includegraphics[width=2.0in]{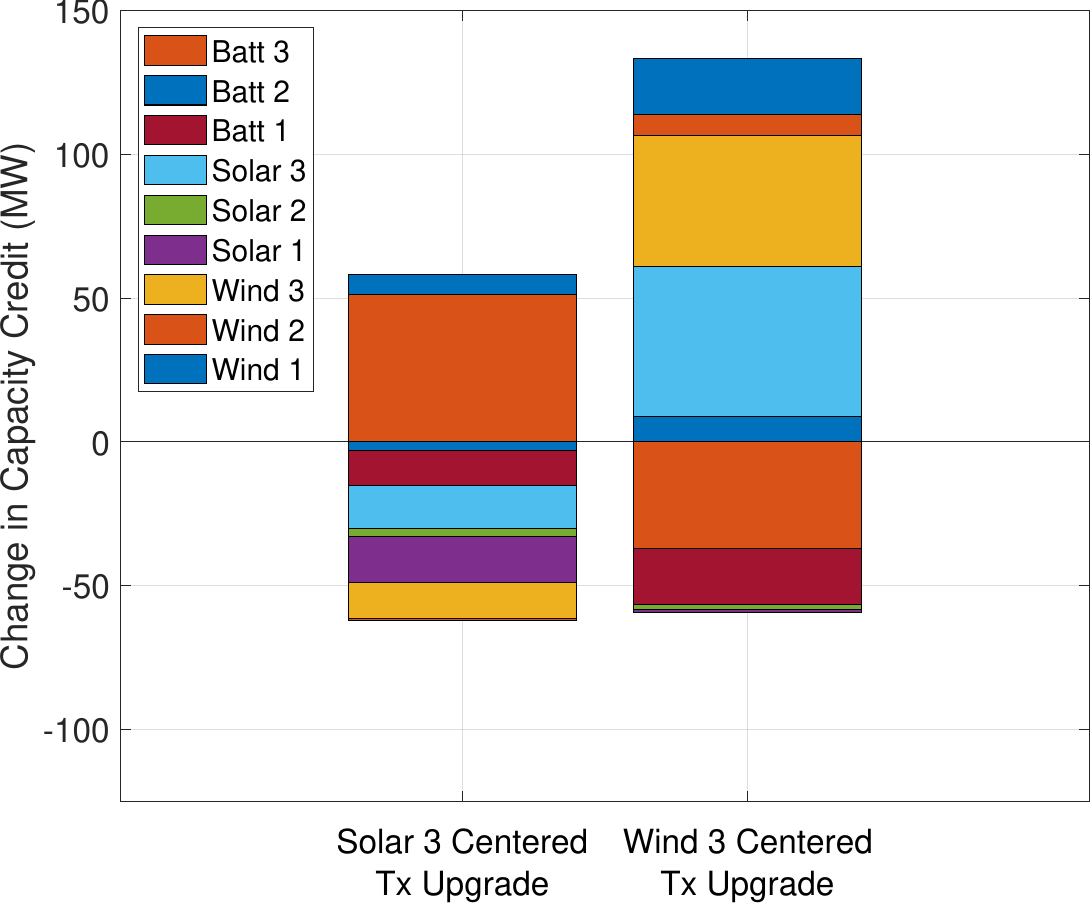}
\vspace{-10pt}
\caption{Change in CCs of the 9-generator portfolio with generator-centered transmission upgrades.}
\vspace{-5pt} 
\label{Gen Trans Upgrade}
\end{figure}

Test 1.1 evaluates the impacts of transmission upgrades centered around Solar Farm 3 and Wind Farm 3 on CCs, as shown in Figure \ref{Gen Trans Upgrade}.  It was expected that when a generator has a low CC due to transmission congestion,
applying upgrades to the associated congested lines would notably boost its CC. However, this is not the case for Solar Farm 3, whose CC instead decreased (-14.9MW) compared to the base case, while BESS 3 experienced the greatest CC benefits from this upgrade. As a result, the total portfolio CC remains approximately the same after transmission upgrades. Conversely, the 2,525MW transmission upgrades around Wind Farm 3 yielded an additional 45.7MW CCs to Wind Farm 3, and the RES-BESS portfolio CC is improved by 74.2MW (+4.0\%). It can be concluded that generator-focused transmission upgrades cannot simply be used to improve the CC of a single generator, and further analyses should be conducted to identify proper transmission upgrade strategies.  

\begin{figure}[h]
\centering
\vspace{-5pt} 
\includegraphics[width=2.5in]{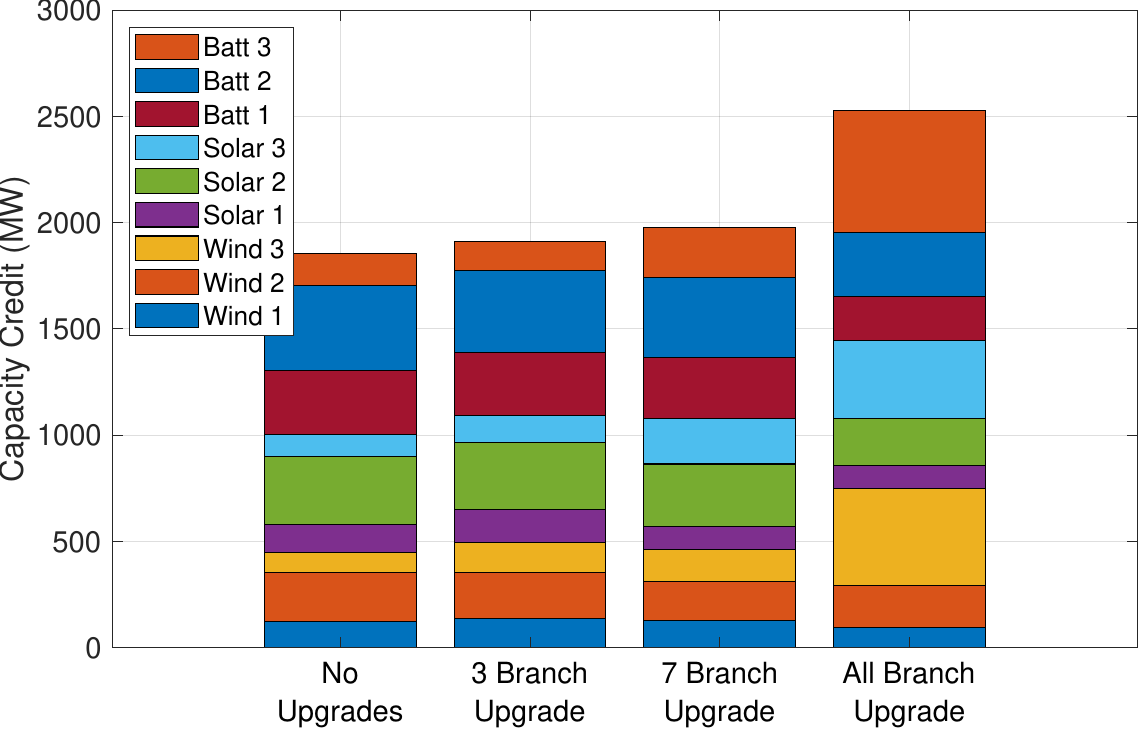}
\vspace{-10pt}
\caption{System-focussed transmission upgrades.}
\vspace{-6pt}
\label{Sys Trans Upgrade}
\end{figure}

Test 1.2 further evaluates the impacts of system-focused transmission upgrades, as shown in Figure \ref{Sys Trans Upgrade}. As the number of transmission branch upgrades increases, the aggregated RES and BESS portfolio CC also increases. In particular, Wind Farm 3, Solar Farm 3, and Battery Farm 3 exhibit notable improvements in their individual CCs while the remaining resources experience a modest reduction in their CCs, indicating reliability contributions are reasonably reallocated toward higher nameplate-capacity generators once transmission bottlenecks are alleviated.

Although the most extensive transmission upgrades yield the highest overall reliability improvement, it is equally important to evaluate the return on capacity (ROC) (i.e., increase in portfolio CCs over transmission capacity upgrades). The “3 Branch Upgrade” adds 825 MW of transmission capacity and raises portfolio CC by 58.5 MW, corresponding to a 7.1\% ROC. The “7 Branch Upgrade” adds 1,925 MW of upgrades and improves the portfolio CC by 125.0 MW (6.5\% ROC). In contrast, the “All Branch Upgrade” applies 63,825 MW of upgrades to achieve 675.7MW portfolio CC improvement, resulting in a substantially lower ROC of 1.1\%. This sensitivity study highlights the impact of transmission constraints in CCs and demonstrates the importance of including transmission constraints in reliability studies. Study results indicate that RTOs can achieve more cost-effective reliability improvements by strategically targeting the most critical transmission bottlenecks rather than pursuing widespread network upgrades that ultimately increase costs passed on to the end consumers.

\subsubsection{Sensitivity Tests of Future Weather Trending Impact}
\begin{figure}[h]
\centering
\vspace{-7pt} 
\includegraphics[width=2.35in]{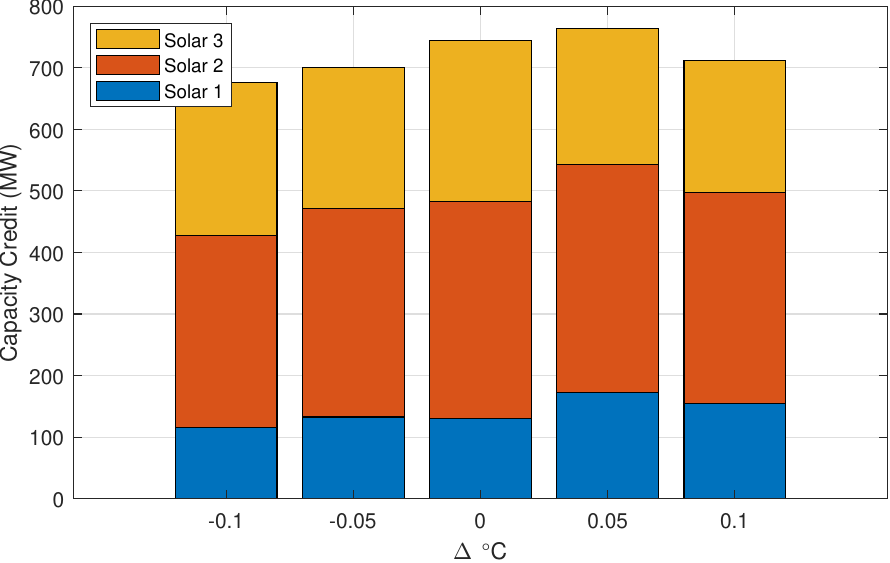}
\vspace{-7pt}
\caption{Future temperature-adjusted CC results.}
\vspace{-5pt}
\label{Temp Sens Analysis}
\end{figure}

 Test 2.1 reevaluates Case 2 across 5 different values of $\beta_\tau$. $\beta_\tau$=0\degree C/yr is used to set the baseline as the current industry practices; $\beta_\tau$=0.05\degree C/yr is the average value across the 12 months for New Jersey,  and $\beta_\tau$=-0.05\degree C/yr is to examine these trends in the opposite direction;  $\beta_\tau$=$\pm$0.1\degree C/yr are further evaluated to account for extremely evolving temperature trends. The impacts of evolving temperature trends on the CC ratings of solar farms are illustrated in Figure \ref{Temp Sens Analysis}. 
Although temperature impacts multiple inputs of the UC problem (e.g., system load, PV performance, transmission AAR) and it is difficult to predict how these factors interact and ultimately drive the CC ratings, a general increasing trend in CC ratings is observed as the evolving temperature increases. This can be associated with the increase in system demand, particularly at peak times, raising the solar farms' reliability contribution. Nevertheless, at $\beta_\tau$=0.1\degree C/yr, the portfolio CC decreases, which could be attributed to the decreased PV performance and transmission derating at high temperatures. Therefore, the non-monotonous impacts of evolving temperature should be incorporated into reliability evaluation, particularly in areas that are experiencing extreme temperature shifts. 

Additionally, Test 2.2 reevaluates Case 1 for multiple values of $\beta_{\text{hurr}}$ ranging from 0 to 1, where $\beta_{\text{hurr}} = 0.0108$ reflecting the historically low increase in hurricane frequency in New Jersey is set as the base case. The sensitivity results indicate that generator CCs exhibit notable changes only when $\beta_{\text{hurr}}$ approaches 1.0, which is nearly two orders of magnitude larger than the base case value. This occurs because hurricane events reduce generator CCs only when they coincide with peak demand. Consequently, even substantially higher hurricane frequencies do not necessarily translate to reduced CCs.

\vspace{-7pt}
\section{Conclusions and Future Work}
This paper studies a new ELCC methodology to enhance CC evaluation for future energy system resources experiencing deep uncertainties. Specifically, the proposed TRACED extends the Delta method, preserving fair CC of individual resources while incorporating locational transmission constraints and evolving weather trends, which have been largely neglected, isolated, or oversimplified in the current industry practice. The TRACED method is evaluated against the marginal ELCC across multiple resource portfolios and operating months. Case studies demonstrate that TRACED provides portfolio-consistent CC allocations by capturing complementary resource interactions and preventing double-counting of shared reliability benefits inherent in the LI marginal ELCC approach, which may otherwise lead to under-procurement of reliability resources. Furthermore, transmission congestion is identified as a major bottleneck in assigned CCs and has a significant influence on the complementary behaviors between resources. Sensitivity studies unveil that targeted transmission reinforcements could substantially enhance system reliability and unlock additional resource values, and that evolving temperature trends are found to impact CCs while hurricane occurrence only impacts CCs when aligned with peak demands.

This work also motivates interesting future research directions. TRACED requires solving LOLEs for both FI and LI systems for each generator of interest, causing scalability challenges for systems with a large generator portfolio. Future work could investigate methods to reduce runtime, such as reusing FI CC results across years when system conditions remain largely unchanged (e.g., no major generator retirements or transmission upgrades). Another area of improvement is the future weather trend model imposed on historical data. This work applies regression methods to capture long-term weather evolution, while other advanced approaches such as statistical models or machine learning could further improve the future weather environment representation for ELCC evaluation.

\end{document}